\documentclass{article}
\usepackage{inputenc}
\usepackage[english]{babel}
\usepackage{verbatim}
\usepackage{graphicx}
\usepackage{caption}
\usepackage{subcaption}
\usepackage{amsmath}
\usepackage{setspace}
\usepackage[ruled,lined]{algorithm2e}
\usepackage{algorithmic}
\usepackage{authblk}
\usepackage[
backend=biber,
style=nature,
sorting=nyt
]{biblatex}
\bibliography{bibliografia}

\title{Measure of gap and inequalities in basic education student’s proficiencies} 
\date{May 2018}

\author[1]{José Francisco Soares}
\author[2]{Erica Castilho Rodrigues}
\author[3]{Victor Maia Senna Delgado }
\affil[2]{Departamento de Estatística, Universiadade Federal de Ouro Preto}
\affil[3]{Departamento de Economia, Universiadade Federal de Ouro Preto}

\begin{document}
\doublespacing
\maketitle

\begin{abstract}
   This study uses students’ performance on standardized tests as evidence of the quality of education and introduces a methodology based on the comparison of performance distributions to produce indicators for both the level achieved by the students and the learning gap between social groups, two inseparable dimensions of quality of education. In the first case, the study compares the distribution of the group observed with a reference distribution, which represents an ideal situation of where students should be. In the second, it compares the performance distribution of students belonging to social groups defined by socioeconomic characteristics. This article uses the Kullback-Leibler divergence to characterize the differences between the distributions. This measure takes into account types of differences not considered by other measures and have solid conceptual justifications. The proposed methodology is used to describe the quality of Brazilian basic education using the test results applied biannually to all Brazilian students of basic education.
\end{abstract}

\section{Introduction}

All countries guarantee their citizens the right to basic education, by law. In order to grant this right, they organize their educational systems. The capacity of these systems to guarantee tangible educational results for students is an indicator of success. 
The first result is to have access to a school, followed by regular attendance and student grade level advancement. Nowadays, another necessary educational result is the acquisition of learnings that students need in their lives. These include, mainly, the basic skills in reading and mathematics.  Although school infrastructure is also essential for positive student results and is, therefore, included in the monitoring of schools, this paper does not consider it.   

For many years, education was analyzed mainly by the number of school grades completed by the students.  This is simple data, but it has allowed the comparison of educational systems among the countries of the world, as shown by Barro and Lee (2013) \cite{barro2013new} and the association of schooling with financial returns. Comparative studies, such as Breen and Jonsson (2005) \cite{breen2005inequality}, have shown the need to consider both the level of the indicators and the inequalities between different social groups in the analyses.  

In recent decades, countries have begun to produce learning data that enable this dimension to be included in the monitoring of educational systems. Learning data is obtained through tests in the native language, mathematics and sciences. These are, nevertheless, not the only areas that deserve attention and in which learning achievement should be monitored.

Several other areas of critical learning are challenging to measure, and their monitoring should be done via their inclusion in the teaching processes. However, there is a lot of information, useful for educational analyses, in the current test results. Several indicators of the level of learning attained by students have been created. The most useful ones, at the international level, take data from PISA, Programme for International Student Assessment.

The monitoring of an educational system with learning data should include at least two categories of indicators. The first category verifies if the students have acquired appropriate knowledge at the level to meet the demands of their lives, referred to in the literature as the quality problem. The second category includes indicators of learning inequalities among students of different socioeconomic groups, referred to as the inequality problem.

The purpose of this paper is to introduce a methodology to deal with inequalities and learning level together. We believe that the term ``quality of education'' should be reserved for analyses that deal with both dimensions, since learning results for only the few should not be evidence of overall educational excellence. 

This paper has four sections, beginning with this introduction. The second section presents the methodology for the study of quality and inequality. Section 3 presents the usefulness of this methodology, using Brazilian data. Finally, section 4 discusses both the methodological and Brazilian results, considering other contributions found in the literature, and the importance of this contribution to the educational debate.

 \section{Metodology}
 
To study educational inequality in student performance, it is natural to consider using the same concepts and methods applied to study income inequality and wealth. This is only appropriate, however, after adaptations.

The study of income inequality assumes that each person has a percentage of the total income of the group. The objective is that all people should have the same income: that is, the income inequality indicators assume that the socially optimal situation is the one in which there is absolute equality of income among individuals. Consequently, any income gap is taken as an expression of inequality, without admitting variation among individuals. This is unexpected, since the ideal of absolute equality is not socially possible in modern democratic societies, which ensure freedom resulting in differences. In addition, the methodology used to measure income inequality assumes that the total income of the group can be, at least in theory, redistributed among the members of the social group being considered.

These hypotheses cannot be used to study educational inequalities. First, in education, a person can increase their learning without decreasing the learning of the other members of the group. That is, the redistribution hypothesis is completely inappropriate in an educational context. In addition, the ideal situation may not be, even utopically, considered as equal learning for all. In a situation where people are not subjected to social constraints, not all of them would choose to acquire the same level of learning in reading, math or science. Therefore, in the educational analysis, the concepts of difference and inequality are distinct.

The solution to the problem of educational inequality is to compare student groups, not individuals. This decision raises two problems. First, it is necessary to define a distribution, in the field of possible performances, which will be referred to in this text as the reference distribution. The educational level indicator will be defined as the distance between this distribution and the empirical distribution. The inequality will be measured by the distance between the distributions of the learnings in the two groups of interest. The second problem is to define how this distance will be calculated.

The definition of the performance reference distribution should be made, ideally, considering the location of the curricular expectations for learning, along the performance measurement scale. This methodology is similar to that used for setting cut-off points to create levels for the pedagogical interpretation of performances, as described, for example, in Cizek and Bunch (2007) \cite{cizek2007standard}. That is, the definition of the reference will always be contextual. This article applies the methodology to Brazilian data, in which the reference distribution was built using a comparative methodology, to be explained in section 3.

To analyze specific hypothesis, it may be necessary to create reference distributions with special features, such as, for example, a distribution in which all students are at the advanced level of performance.

 \subsection{Distance between distributions}

To create an indicator of the performance level achieved as well as to quantify the inequality among groups of students, it is necessary to define a way to measure the distance between distributions of learnings. However, the measure to be used in this paper to characterize how far one distribution is from another does not satisfy all the conditions of a distance in the mathematical sense. Although this text uses the term distance, to show the misuse of language, the word distance will always appear in italics.

Let $f(y)$ and $f_0(y)$ be density functions with the same domain, whose distance is to be measured. The ratio $f(y)/f_0(y)$, measured at each point of the domain, is an expression of the point difference between the distributions. In this case, the equality of the densities would occur when this ratio is equal to 1 at all points in the domain. These point indicators of the difference among distributions should be aggregated to obtain a synthetic  measure of the distance between the distributions. 

\subsection{Relative Distribution}

The concept of relative distribution, dealt with in the social sciences by Handcock and Morris (1998) \cite{handcock1998relative}, creates a theoretical framework that facilitates the choice of the proper way of aggregating these ratios in a measurement.

Defining $F(y)$ and $F_0(y)$ as the cumulative distribution functions of two distributions, the relative distribution is defined as:

$$
R=F_0(Y)
$$
where $R$ is a random variable defined in the interval $[0, 1]$. The relative distribution represents the rank that an observation generated from a distribution with $f(y)$ density occupies at the $f_0(y)$ density. If we define $f(y)$ as the density function of the analyzed students’ proficiency and $f_0(y)$ as the density of the reference distribution, the relative distribution is nothing more than the rank of each student in the reference distribution.

The cumulative distribution function of the relative distribution is given by:

\begin{equation}\label{eq:cdf}
G(r) = F(F^{-1}_0(r)) \mbox{ for } 0 \leq r \leq 1
\end{equation}
where $r$ is a proportion, $F^{-1}_0(r)=\inf\{y \mid 
F_0(y) \geq r\}$ is the quantile of the distribution $F_0(y)$. The density function of $R$ is given by:

\begin{equation}\label{eq:densidaderelativa}
g(r) = \frac{f(F_0^{-1}(r))}{f_0(F_0^{-1}(r))}\mbox{ for } 0 \leq r \leq 1
\end{equation}

Ali and Silvey (1966) \cite{ali1966general} showed that the relative distribution is a sufficient statistic for the comparison of two densities. Therefore, the search for a definition of distance between two distributions can be reduced to the syntheses of the relative distribution.

If the two distributions being compared are equal, their relative distribution is a uniform variable and $G(r)$ is a straight line with a $45^{\circ}$ slope.

\begin{figure}
\begin{subfigure}{.5\textwidth}
  \centering
  \includegraphics[width=.8\linewidth]{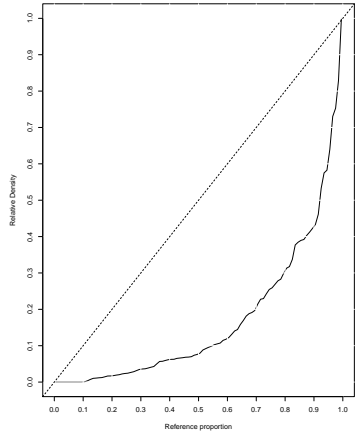}
  \caption{}
  \label{fig:sfig1}
\end{subfigure}%
\begin{subfigure}{.5\textwidth}
  \centering
  \includegraphics[width=.8\linewidth]{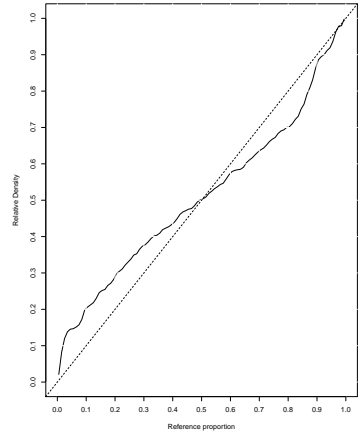}
  \caption{}
  \label{fig:sfig2}
\end{subfigure}
  \centering

\begin{subfigure}{.5\textwidth}
  \centering
  \includegraphics[width=.8\linewidth]{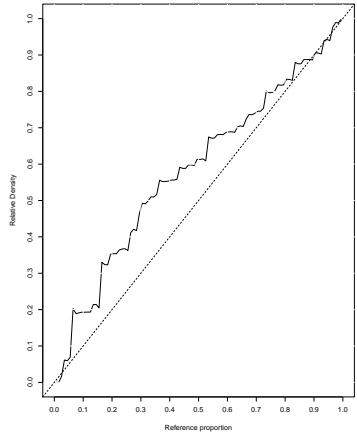}
  \caption{}
  \label{fig:sfig2}
\end{subfigure}
\caption{Examples of configurations of the cumulative distribution function $G(r)$ for the case where the observed distribution is worse than reference (a), have both behaviors for the same group (b) and better than reference (c).}
\label{fig:graficos_cdf}
\end{figure}

The graphic expression of the cumulative distribution function, defined in (\ref{eq:cdf}), facilitates the understanding of the difference between the distributions, which we want to characterize and measure. Figure \ref{fig:graficos_cdf} gives some examples. Graph (a) shows educational situations in which the distribution of the students’ proficiencies of the group of interest is below the reference distribution, which is the most common situation. Graph (b) shows a situation in which the distribution of the proficiencies of the group of interest dominates the reference distribution at some intervals. Graph (c) shows an uncommon situation, in which the distribution observed exceeds the reference. If this occurs frequently, it should be evaluated whether the reference distribution was chosen properly.

 \subsection{Kullback-Leibler \textit{distance}}

The area between $F(y)$ and $F_0(y)$ is the most obvious option for defining the distance between the two distributions. However, this area is simply the difference between the means of the two distributions. This can easily be seen using the following result:

$$
E(Y)=\int_{0}^{\infty} (1- F(y))dy 
$$

Therefore, this is a limited option for measuring the distance between two distributions, because it considers only one kind of difference between them. One alternative, used by Soares and Marota (2009) \cite{soares2009desigualdades}, is to take the area between the relative cumulative distribution function, shown in equation (\ref{eq:cdf}), and the straight line, as shown in the graphs in Figure  \ref{fig:graficos_cdf}. Again, this option is limited to taking a mean as the distance, although, in this case, it is a mean of the ranks. This result is a consequence of the fact that

$$
E(R)=\int_{0}^{1} (1- F(r))dr
$$
and, therefore, the area between $F(r)$ and the straight line is nothing more than a linear function of $E(R)$.

This average presents a gain compared to the previous one, since it is not affected by atypical observations. However, it still does not use all the available information in order to compare the two distributions.

Given this, it is necessary to seek out another way to define the distance that allows considering all types of differences among the distributions. Naturally, each type of difference has educational significance and demands different educational policies in order to be overcome.

Since the relative distribution provides all the information necessary for the comparison of two distributions, a distance measurement should be calculated with a density synthesis $g(r)$. Handcock and Morris (1998) \cite{handcock1998relative} suggest the following class of functions:

\begin{equation}\label{eq:phi}
D_\phi(F,F_0) = \int_0^1\phi \left(  g(r) \right)dr
\end{equation}
where $\phi$ is any continuous and convex function, defined in the interval $(0, 1)$. Defining $\phi$ as:

$$
\phi(g(r)) = g(r) \log\left(g(r)\right)
$$

the expression in (\ref{eq:phi}) is equal to

$$
D_{KL}(F,F_0) = \int_0^1 g(r)\log(g(r))dr
$$

replacing the expression of $g(r)$, given in (\ref{eq:densidaderelativa}), we have:

$$
D_{KL}(F,F_0) = \int_0^1 \frac{f(Q_0(r))}{f_0(Q_0(r))}\log\left(\frac{f(Q_0(r))}{f_0(Q_0(r))}\right)dr
$$

Changing variables, $y=Q_0(r)$, we have $r=F_0(y)$ and $dr=f_0(y)dy$. Therefore, we obtain:

$$
D_{KL}(F,F_0)=\int_0^1 \frac{f(y)}{f_0(y)}\log\left(\frac{f(y)}{f_0(y)}\right)f_0(y)dy=\int_0^1 f(y)\log\left(\frac{f(y)}{f_0(y)}\right)dy.
$$

This is the expression of the Kullback-Leibler divergence, henceforth represented by (KL) (\cite{kullback1951information}, \cite{kullback1997information}) between $f(y)$ and $f_0(y)$. This will be the measure of the distances between the distributions used in this study. This choice is justified based both on the concept of entropy, central in information theory, introduced by Shannon (1948) \cite{shannon1948mathematical}, and on the concept of likelihood.

\subsection{Entropy and Redundancy}

Entropy is a central concept in information theory, where a basic problem is to measure the information obtained as an event occurs. The mathematical expression of this measure is obtained as a consequence of the following postulates.

\begin{itemize}
    \item $I(p)$ is decreasing in $p$, as the greater the probability of the event, the less information it carries;
    \item $I(1) = 0$, a certain event has no information;
    \item $I(p)\geq 0$, the entropy is a positive number;
    \item $I(p_1p_2) = I(p_1) + I(p_2)$, the entropy of independent events is the sum of the individual entropies.
\end{itemize}

With these hypotheses, it can be shown that the only way to measure the information is to use the expression

\begin{equation}\label{eq:entropia}
I(p)=\log(1/p)=-\log(p)
\end{equation}
therefore, the mean amount of information of a system with $n$ events is given by:

$$
E\left(I(p)\right) = -\sum_{i=1}^n p_i \log(p_i)
$$
where $n$ is the number of possible outcomes and $p_i$ is the probabilities associated with each of them.

A system with maximum entropy is that in which all events have the same probability since, in this case, it is more difficult to predict the result before it occurs. From this arises the concept of redundancy, the difference between the maximum entropy and the entropy that is actually observed. The redundancy quantifies the reduction of uncertainty that a probability distribution presents, relative to the situation of maximum uncertainty.

The concepts of entropy and redundancy may be used to study income inequality among members of a social group. To do so, it is sufficient to consider a measure of inequality as the difference of the entropy of a system where each individual has a given portion of the total income, and the situation in which everyone has the same income. This is the definition of the Theil Index \cite{theil1967economics}. That is, this indicator measure how much it is necessary to change the observed income distribution in order to reach the ideal of equality.

Its expression is given by:

$$
T = -\sum_{i=1}^n \frac{1}{n}\log(1/n)+\sum_{i=1}^n p_i\log(p_i)
$$
where $n$, in this case, is the number of individuals of the population and $p_i$ is the portion of the income that each one receives. The index may be rewritten as:
$$
T = \log(n)+\sum_i p_i\log(p_i) = \log(n)\sum_i p_i + \sum_i p_i\log(p_i)
$$

putting $p_i$ in evidence::

\begin{equation}\label{eq:theil}
T = \sum_i p_i\Big[\log(np_i)\Big] = \sum_i p_i \log\left(\frac{p_i}{1/n}\right)
\end{equation}

Designating $x_i$ as the income of the i-th individual and pi as the proportion of the total income that he/she receives, the equation becomes:

$$
T=\sum_i \frac{x_i}{\sum_i x_i} \log\left(\frac{x_i}{\sum_i x_i /n}\right)=\frac{1}{n}\sum_i \frac{x_i}{\mu} \log \left(\frac{x_i}{\mu}\right)
$$
where $\mu$ is the average income of the population. This last expression is the best known form of the Theil index.

The expression obtained in  (\ref{eq:theil}):

$$
T=\sum_i p_i \log\left(\frac{p_i}{1/n}\right)
$$
is the \textit{distance}, KL, between two distributions. This is not defined, however, in the domain of incomes. Rather, it is defined in the domain of persons who compose the social group in which the inequality is studied. In the first distribution, the density of each person is equal to their proportion of the total income and, in the second distribution, all persons have the same proportion of income, therefore, the densities at all points are equal.

Therefore, taking the two sets of proportions as the densities of two distributions in the domain of the persons of the social group being considered, the problem of defining the distance between those distributions may be treated as a problem of calculating the redundancy between the two situations. As the entropy is uniquely defined by expression (\ref{eq:entropia}), the redundancy is also uniquely defined and, as shown, is the KL between the distributions. Thus, accepting the postulates that generated the entropy, any other way of measuring the distance between the distributions will produce a non-optimal value.

\subsection{KL as a measure of likelihood}

The use of the Kullback-Leibler \textit{distance} is also justified for likelihood, either for the construction of a hypothesis test or as an expression of the likelihood of the observed data.

Shlens (2014) \cite{shlens2014notes}  shows that KL is a measure of how likely the model $f(y)$ is, if the data were generated from $f_0(y)$.

The same idea may be shown through a form of hypothesis testing. In order to test the hypothesis that the data observed in the distribution of the group of interest were, in fact, generated by the distribution from the reference distribution by the likelihood ratio, the ratio of the densities $f(y)/f_0(y)$ should be used. Under the null hypothesis, the data come from the distribution of the reference $f_0(y)$  and, under the alternative hypothesis, the data come from the distribution of interest  $f(y)$. Therefore, the statistic of the test of the ratio of likelihood, $L$, is given by:

\begin{equation}\label{eq:testerazao}
L = -2 \log\left(\frac{f_0(y)}{f(y)}\right)=2 \log\left(\frac{f(y)}{f_0(y)}\right)
\end{equation}

If the alternative hypothesis is true, the data come from the distribution $f(y)$. In this situation, the test statistic expectation presented in (\ref{eq:testerazao}) is:

$$
E(L) = 2 \int_{-\infty}^{\infty}  
 \log\left(\frac{f(y)}{f_0(y)}\right)f(y)dy
$$

This is exactly the Kullback-Leibler divergence between the two distributions. Therefore, KL can be seen as the expected value of the log-likelihood ratio when the data come from the distribution established in the alternative hypothesis. The larger this quantity, the more evidence we have that the data come from the $f(y)$ distribution. When the data are generated from the $f(y)$ distribution, KL measures how much more likely the model is, in relation to  $f_0(y)$. For more details on this discussion, \cite{eguchi2006interpreting} can be consulted.

These observations and associations show that KL is a technically solid and highly intuitive option to measure the distance between the two distributions, in the context which interests us.

An important point to stress is that this distance has a disadvantage, as it presents only positive values. This prevents us from knowing if the observed curve is above or below the reference. Situations in which the distribution of student performance is better than the reference are rare, but when the municipalities and subgroups within them are analyzed they do appear. To solve this problem, the sign of the difference of areas between the cumulative distribution function defined in (\ref{eq:cdf}) and the $45^{\circ}$ straight line will be used.

\section{Study of intra-school exclusion in Brazil} 

Basic education students are subjected to intra-school exclusion when they have not learned what they need to learn at a level that allows them to function in society; or, who have learned less than students from another social group of reference. In both situations, these students will have their access to social opportunities limited, either to advance in their studies or to have access to jobs that provide greater satisfaction and compensation. Although these two types of educational exclusion are socially important, the Brazilian educational debate has focused only on the first dimension, producing very useful information that has been widely used for educational planning but that does not describe the educational situation completely.

The aim of this section is to show how KL can be used to produce measures that characterize the two dimensions of school exclusion and, thus, provide visibility concomitantly to both. In this section, intra-school exclusion is studied in the Brazilian municipalities, using all the data from the Prova Brazil between 2007 and 2015. It is, therefore, an historical characterization of the municipality and does not deal with the evolution of the respective educational situation. Additionally, we present the results only for mathematics in the fifth year and the only inequality considered is the socioeconomic one.
\subsection{Data base}

The data used in this article were obtained within the scope of the SAEB (Basic Education Assessment System), under the responsibility of the INEP (National Institute of Educational Studies and Research Anísio Teixeira), a federal agency linked to the Ministry of Education (MEC) that is responsible for the assessment of both Basic and Higher education. The main objective of SAEB is to measure the performance of Brazilian students in elementary education schools, as well as to collect information about the students, their teachers and the schools they attend. This information can be used to construct a diagnosis of Brazilian basic education and provide information that supports the formulation and monitoring of educational policies.

Today, the Saeb comprises three large-scale external evaluations. The first is the National Assessment of School Performance (Anresc), known as the Prova Brasil, created with the aim of assessing, through census, student learning in the 5th and 9th years of elementary education, and in the third (final) year of secondary education, taught in schools of the public school system. The second is the National Evaluation of Basic Education (Aneb), which tests a sample of public and private schools in the same school years of the Prova Brasil. The third is the National Literacy Assessment (ANA), whose aim is to verify, the level of literacy of students in the third (final) year of public school basic education. For a review of the literature on the initial cycles and on the implementation of the SAEB, see Bonamino and Franco (1999)  \cite{bonamino1999avaliaccao} and Gatti (2009) \cite{gatti2009avaliaccao}.

In the Prova Brasil, the performance of the students is estimated using a three-parameter Model of Item Response Theory (IRT), and the resulting scores are given the technical name of proficiency (KLEIN, 2009) \cite{klein2009utilizaccao}. The proficiency scale, originally in standard deviations, is transformed to values from 0 to 500 points. As the proficiency scale is the same for the different editions of the Prova Brasil, due to the equalization process used to analyze the tests, the variations in the proficiency of students of a given school year, among the editions, reflect the improvement or worsening in the learning of the evaluated cohorts. In addition, since the same scale is used to express student scores for different school years or grades, 5th year students typically have lower proficiencies than 9th year students.

The versions available on the INEP website were used for this article. They were processed using the R software \cite{citeR}, with the available codes.

The socioeconomic status (SES) of the students was calculated using the same model of Item Response Theory used by Alves and Soares (2015) \cite{alves2015indice}. It is important to highlight that only data from the contextual questionnaires of the Prova Brasil, during the period examined, were considered for this calculation. However, the results obtained for each school have a high correlation with the SES indicator for the school, adopted by INEP, that uses data from other evaluations. To measure the level of student learning in each municipality, the KL distance between the learning distribution of all students in Brazilian municipalities and the reference distribution was calculated. This was constructed as explained in the next section of this article. However, only municipalities with more than 100 students in all cycles of the Prova Brasil are included in the analysis.

\subsection{Reference Distribution}

The establishment of a reference distribution requires the definition of a distribution that is taken as a goal for educational policy planning. The research that would allow the construction of the reference distribution only by pedagogical criteria has not yet been completely done in Brazil.

Therefore, we will use a reference constructed using some assumptions that may also be objects for further analysis in future articles. In favor of the reference used, we highlight that it meets the criteria for good levels of proficiency in the intervals proposed by Soares (2009) \cite{soares2009indice}, and such a distribution may be considered desirable. It should be noted that nothing prevents other references from being used, but we will present the justifications for what has been done here.

The reference distribution used in this article is made in a comparative manner. The participation of Brazilian students in the PISA allows the construction of the comparison distribution, as done by Soares et. al. (2010) \cite{soares2010measuring}. These authors created a \emph{typical country}, consisting of a combination of the following countries: Australia, Austria, Belgium, Canada, Denmark, England, Finland, France, Germany, Holland, Iceland, Ireland, Italy, Japan, Korea, New Zealand, Norway, Portugal, Spain, Sweden, Switzerland and the United States.

For each one of these countries, using the sample weight of each student, the values of the 100 percentiles of the proficiency distribution in the PISA were calculated. The data from 2000 were used for Reading, and the data from 2003 were used for Mathematics. These are the cycles in which the respective scales were defined. Each percentile of the distribution of the typical country was defined as the mean of the percentiles of each selected country. This value will be denoted by $Z_r$, where $r$ assumes integer values between 0 and 100, and indicates the distribution percentile. In parallel, the 100 percentiles of the proficiency distribution of Brazilian students in Reading and Mathematics in the same PISA cycles were also obtained, and are denoted by $X_r$.

As students from the \textit{typical country} perform better than the Brazilian students, there is a positive difference between the percentiles of this ``country'' and the respective percentiles of the Brazilian students in the same PISA test. This distance, $\delta_r$, expressed in units of the standard deviation of the proficiency distribution of the Brazilian students on the PISA ($\sigma$), is given by:

$$
\delta_r=\frac{Z_r-X_r}{\sigma}
$$

The $\delta_r$ value indicates how many standard deviations each distribution percentile of Brazilian students in the PISA should increase, so that the performance of the Brazilian students is equal to the percentiles of the \textit{typical country}. Another challenge, after this step is established, is to consider what the reference of the \emph{typical country} would be on the SAEB scale. This challenge is tantamount to the hypothetical situation in which those countries take the Prova Brasil. We will comment on the assumptions, related to this choice, at the end of this subsection.

To obtain the percentiles of the reference distribution on the SAEB scale, the translation $\delta_r$ was applied to each distribution percentile of the 1997 (year of definition of the scale) SAEB. That construction is detailed, above. The percentiles of the reference distribution on the SAEB scale ($Y_r$) are obtained from the following expression:

$$
Y_r = Y_r' +\delta_r s 
$$

In this expression, $Y_r$ is the value on the SAEB scale of the percentile $r$ of the reference distribution, and $Y_r'$ is the same percentile before translation, $s$ is the standard deviation of the SAEB distribution from 1997. Table \ref{tab:referencia} synthesizes and illustrates the calculations for a set of percentiles, chosen to illustrate the procedure. The calculations are presented for the 9th year, due to their closeness in age, 15 years, to the students evaluated in PISA. The $\delta_r$ is maintained for the 5th year, but $Y_r'$ is replaced by the test values for that year. 
\begin{table}[h]
\centering
\caption{Construction of the reference distribution in the SAEB metric for students in the ninth year - Mathematics - of the Brazilian elementary school}
\label{tab:referencia}
\begin{tabular}{llll}
\hline\hline
Percentil & Percentile values of & Translation necessary in   & Percentiles of the reference \\ 
& the 1997 SAEB distribution & Standard Deviations & distribution\\
\hline\hline
5 & 173 & 1.55 & 240 \\ \hline
15 & 197 & 1.64 & 268 \\ \hline
30 & 220 & 1.72 & 295 \\ \hline
50 & 248 & 1.75 & 324 \\ \hline
75 & 283 & 1.69 & 356 \\ \hline
90 & 317 & 1.53 & 383 \\ \hline
95 & 338 & 1.35 & 396 \\ \hline\hline
\end{tabular}
\caption*{Source: Soares e Delgado (2016) \cite{soares2016medida}.}
\end{table}

Knowing the 100 percentiles $r$ of the reference distribution on the SAEB scale, it is possible to generate a sample of the distribution. The usual procedure for generating a sample of a random variable $Y$ uses the fact that, if $F$ is its cumulative distribution function, the $F(Y)$ transformation has a uniform distribution defined in the interval $[0, 1]$.

To generate a sample of $Y$, one need only generate a value $u$ of a $U(0, 1)$ and then take $y=F^{-1}(u)$. However, there is no analytically defined function, only the percentiles of an empirical distribution. This means that $y=F^{-1}(u)$ is only well defined for integers between 0 and 99.

Therefore, the central idea is to generate a random integer $u$ between 0 and 99. This number will define between which percentiles the sample should be generated. Observation $y$ is, then, generated from a continuous uniform between these two values.
The following algorithm describes how this simulation is produced.

\begin{algorithm}[H]

\begin{enumerate}
    \item Generate an integer $u$ between 0 and 99;
    \item Find the boundary values of $F^{-1}(u)$ and $F^{-1}(u+1)$;
    \item Generate an observation $y$ from a uniform in the interval $\left(F^{-1}(u),F^{-1}(u+1)\right)$.

\end{enumerate}

\caption{How to generate a sample of a distribution from its percentiles.}

\end{algorithm}

\begin{figure}
  \centering
  \includegraphics[width=1\linewidth]{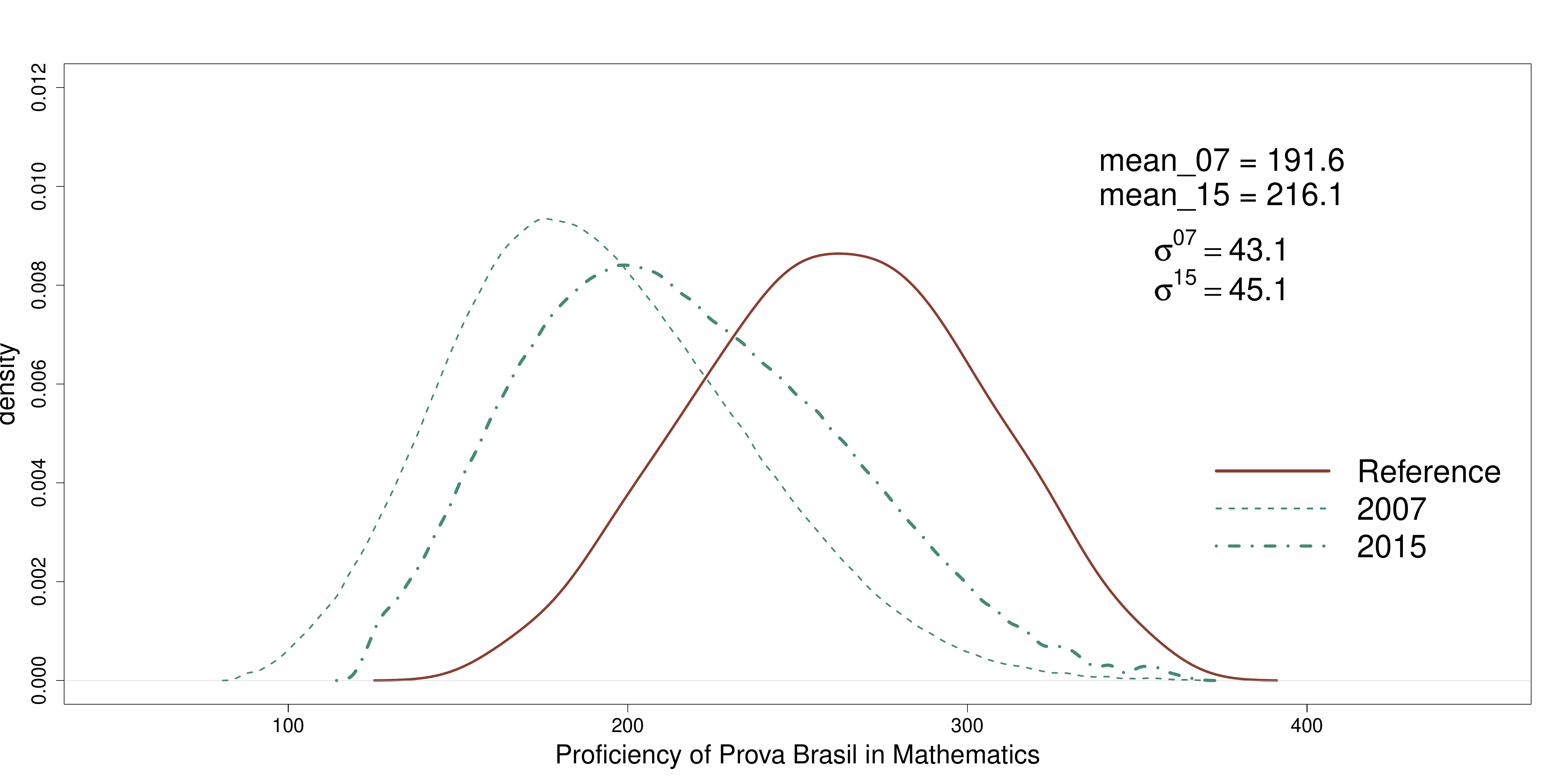}
\caption{Observed distributions and references, 5th Year Mathematics, 2007-2015.}
\label{fig:graficos_referencia}
\end{figure}

Figure \ref{fig:graficos_referencia} shows an example of a distribution generated in this manner, which is the indicated reference.

It is important to emphasize three assumptions regarding the construction of the reference distribution. The first, the strongest, is that the matrix of the two tests are comparable. It is known, however, that the SAEB reference matrix is different from the PISA knowledge matrix. The knowledge and skills measured in the two tests are different \cite{soares2013evoluccao}. One consequence of having this broken assumption is that, even if the SAEB reference distribution is reached, there would be no guarantee that the performance of Brazilian students on the PISA would be equivalent to the performance of the students of the \textit{typical country}.

The second assumption is that, even if the first assumption does not remain true, the fact that we obtain the value of $\delta_r$ for students of similar age, 15 years, the age of the target audience of students selected for the international test, should indicate an array of competencies to be developed that align with minimum international requirements. There is no reason for Brazil to be left behind in this matter. Some compatibility with minimum standards of international excellence should be sought.

In spite of this, it is known that a relevant difference occurs in the sample space because PISA prioritizes by age and the SAEB evaluates by year of schooling. For students who have not been retained, the typical age of the 9th year of school is 14 to 15 years, the same as PISA. Even though the picture has improved in recent decades, Brazil still has a meaningful number of students in grade-level age distortion; that is, students older than 15 years and attending the 9th year of school.

The third assumption is that both tests have some methodological compatibility regarding Item Response Theory, as they both have three parameters. Regarding PISA, Aguiar (2010) \cite{aguiar2010funcionamento} shows how the same item can work differently between Brazilian and Portuguese students, as some are easier for Brazilians and vice versa. Soares and Nascimento (2013) \cite{soares2013evoluccao} emphasize that the SAEB is more content-oriented and PISA is more focused on applications of the Brazilian curriculum.

The assumptions behind the reference application are demanding. In defense thereof, we can highlight the urgent need for Brazilian teaching and learning goals. This is even more urgent, when compared to international quality. Other references can be used; the present study need not be the only possible reference. One idea would be to have students from Portugal (a country that has better PISA results than Brazil) take the SAEB exams. If this existed, such a performance distribution would also serve as a possible reference. However, none of this rules out the need for further studies to verify the adequacy of the assumptions made here.

The inequality is measured by calculating the KL between the performance distributions of the students of the two groups being considered. Here, they are restricted to the groups of higher and lower socioeconomic levels. This calculation included only municipalities in which more than 50\% answered the contextual questionnaire of the Prova Brasil and, in addition, in which each of the groups involved has a minimum of 20 students. These precautions seek to ensure that the value of the measures presented is not idiosyncratic.

As with any statistical measurement, the nominal values of KL do not automatically have a substantive interpretation. One way to construct such an interpretation is to divide its domain into some intervals and to assign a substantive value to each. Those values would indicate if the respective values are pedagogically relevant and should impact school policies, or if they are small and represent only natural fluctuations between years and groups.

The scales used to measure performance are interpreted by creating ways of allocating each student in one of four levels: Below Basic, Basic, Appropriate and Advanced. The form of definition and the chosen labels, for those groups, have clear pedagogical contents. Thus, to construct a substantive interpretation for the KL measure, it is sufficient to create an extension of the interpretation of individual measures that allows values to be given to the distributions of performance in groups of students.

For the present article, we chose to divide the KL values into five groups, for the purpose of interpretation, and to identify them: Low, Medium-Low, Medium, Medium-High and High. For example, the KL values corresponding to the Low range reflect situations in which the student performance distribution is far from the reference. Therefore, the problem now is to identify which distributions will receive each of these labels. This was done using a two-stage methodology.

In the first stage, for each municipality and each year of the Prova Brasil, the profile of each municipality was calculated. That is, the vector of their percentages in the Below Basic, Basic, Appropriate and Advanced levels. Using the K-means methodology to create conglomerates, these profiles were aggregated into five groups. The centroids of these groups correspond to percentages typical of each interval.

The second stage of the algorithm consists of allocating each municipality using the Euclidean distance between the vector of the four values for each municipality, in each year, for the group defined by the centroids obtained in the previous step. Concomitantly, the value of the
respective KL, to the reference distribution, was registered. The steps of this algorithm are detailed, below.

\begin{algorithm}[H]

\begin{enumerate}
    \item Calculate the proportion of students in each of the learning levels (Below Basic, Basic, Appropriate and Advanced) for each municipality;
    \item Apply a 4-dimension vector with each of these locations;
    \item Apply the k-means method to these vectors by defining the number of groups to be equal to 5;
    \item Calculate the centroid of each of these 5 groups;
    \item Find the closest centroid, using the Euclidian distance as the metric, for each municipality;
    \item Attribute the classification to the reference municipality closest to the centroid;
    \item After allocating all municipalities to the 5 groups, the KL cut-off points will be the 95\% percentile of each of these groups.
\end{enumerate}

\caption{How to find the interpretive limits of KL.}

\end{algorithm}

This process generated the KL cut-off points shown in Table \ref{tab:cortes_KL}.

\begin{table}[]
\centering
\caption{Cut-off points defining KL measurement levels.}
\begin{tabular}{l|l|l}
\hline\hline
KL Level    & Lower Limit & Upper Limit\\ \hline\hline
Low       & -4                   & -1.7                 \\ \hline
Medium-Low & -1.7                 & -1.1                 \\ \hline
Medium       & -1.1                 & -0.7                 \\ \hline
Medium-High & -0.7                 & -0.4                 \\ \hline
High        & $> -0.4$                 &                     \\ \hline\hline
\end{tabular}
\caption*{Source: Authors.}
\label{tab:cortes_KL}
\end{table}

\begin{figure}[h]
  \centering
  \includegraphics[width=1\linewidth]{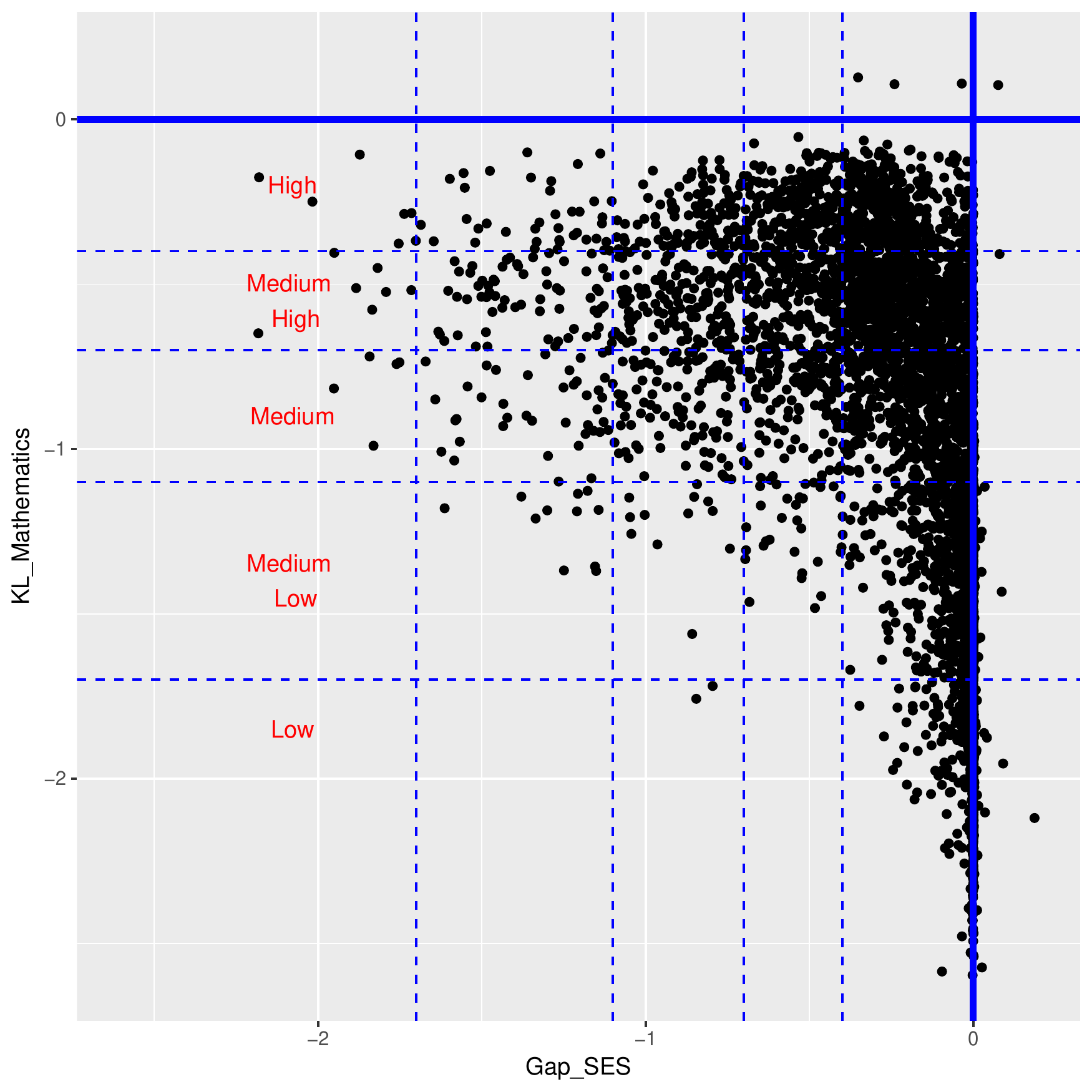}
\caption{Graph of the SES vs KL Gap for all Brazilian municipalities.}
\label{fig:hiato_NSE}
\end{figure}

Figure \ref{fig:hiato_NSE} presents a graph in which the vertical axis shows the KL values, representing the educational level. The horizontal axis shows the gap between the high and low SES groups; that is, the KL between the distribution of the low SES and high SES groups. The high SES group is taken as a reference, here. This means that negative values for this measurement refer to those municipalities in which the group having the highest socioeconomic level also has the highest educational level. Positive values represent the inverse situation. The solid, continuous line represents the mean of the KL for all municipalities, and the dashed lines are the limits shown in Table \ref{tab:cortes_KL}. The same information is shown in Table \ref{tab:kl_hiato_nse}.

\begin{table}[]
\centering
\caption{Measures of Level and Inequality in Brazilian Municipalities}
\label{tab:kl_hiato_nse}
\begin{tabular}{l|l|l|l|l|l|l}
\hline\hline
            & High (Gap) & Medium High & Medium & Medium Low & Low & Total \\ \hline
High (KL)   & 8             & 40          & 111   & 206        & 478  & 843   \\ \hline
Medium High  & 8             & 65          & 146   & 223        & 701  & 1.143 \\ \hline
Medium       & 5             & 42          & 104   & 143        & 743  & 1.037 \\ \hline
Medium Low & 0             & 11          & 12    & 31         & 995  & 1.049 \\ \hline
Low      & 0             & 0           & 2     & 0          & 568  & 570   \\ \hline\hline
\end{tabular}
\end{table}

\begin{figure}[h]
  \centering
  \includegraphics[width=1\linewidth]{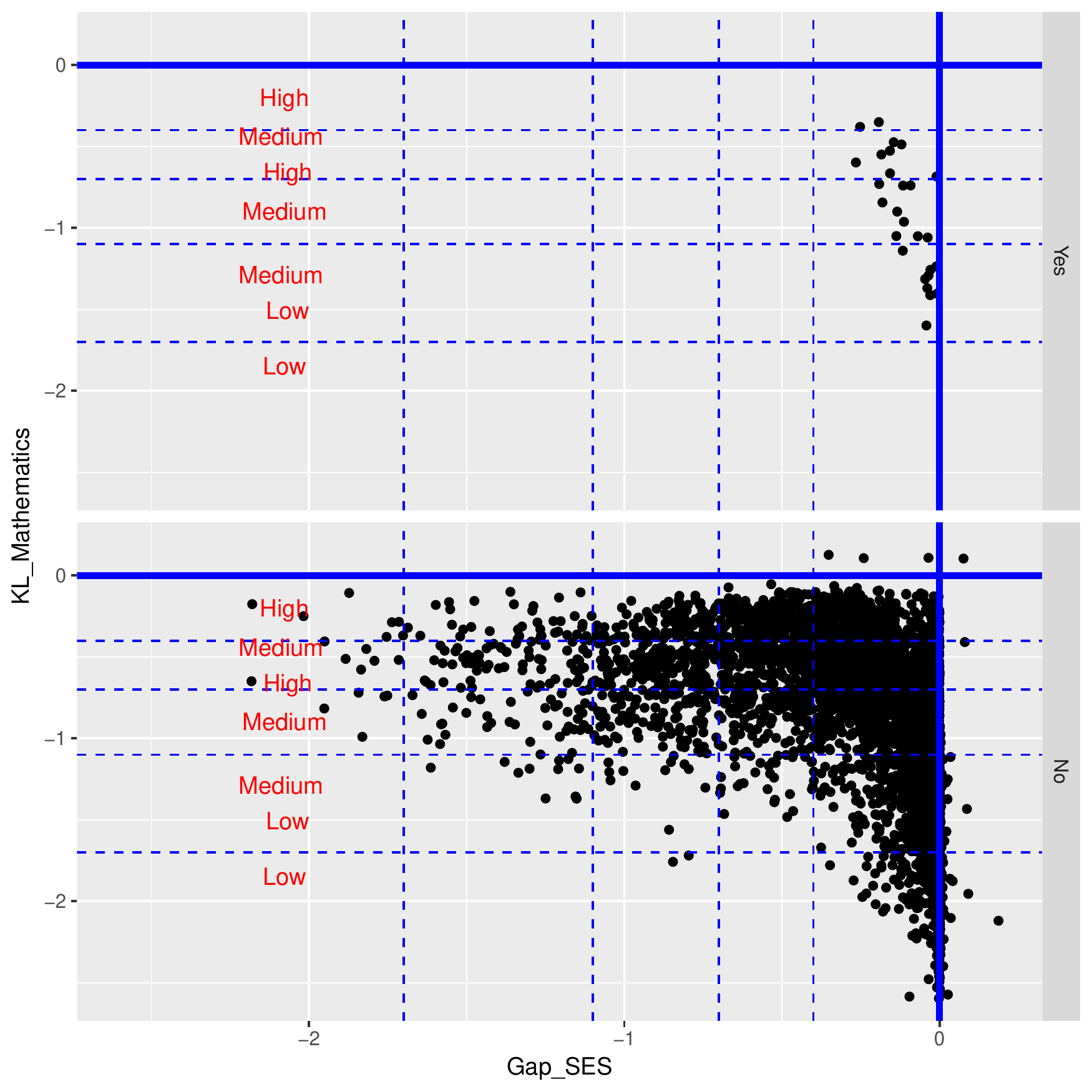}
\caption{Graph of the SES Gap vs KL with the capitals highlighted.}
\label{fig:hiato_NSEcapitai_destaque}
\end{figure}

\begin{figure}[h]
  \centering
  \includegraphics[width=1\linewidth]{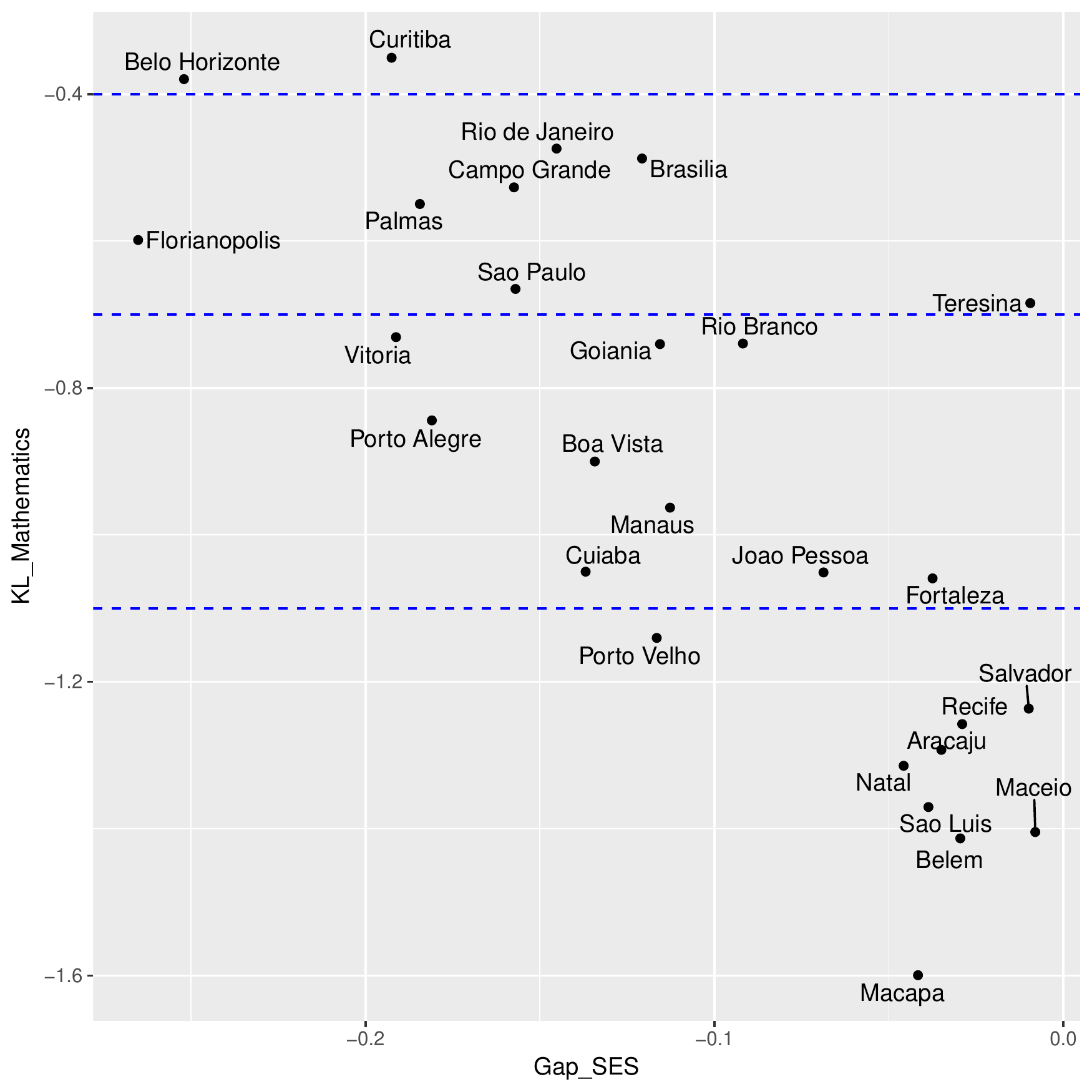}
\caption{Graph of the SES Gap vs KL for capitals.}
\label{fig:hiato_NSEcapitais}
\end{figure}

Figure \ref{fig:hiato_NSEcapitai_destaque} presents the Brazilian capitals, individually. Figure \ref{fig:hiato_NSEcapitais} presents a zoom for this set of municipalities. All show a negative gap, indicating that students with low SES learn less than those with high SES. It is observed that Salvador and Teresina have lower inequalities associated with the SES, but Teresina has a clearly better situation in terms of the level of learning obtained.

\section{Discussion}

This article produced indicators for the two instances of intraschool exclusion: the difference between the performance of students from different social groups - the problem of inequality - and the distance of the performance distribution of a group of students to the expected situation - the problem of quality. The usual model for measuring income inequality among social groups, described in Atkinson (1983) \cite{atkinson1983economics}, can be expressed as the comparison of two distributions whose common domain consists of the people in a population. The first is the distribution taken as ideal, in this case fixed as a discrete uniform distribution. The second is the empirical distribution of the incomes of the people of the social group under analysis. The Theil coefficient, which is actually the measure of the redundancy of the empirical income distribution and, mathematically, the measure of Kullback-Leibler divergence between the two distributions, can be used to measure the distance between these distributions.

The present article, to deal with the educational inequalities, did not formulate the problem of learning inequality in the domain of people, as done in the case of income distribution, but rather in the domain of proficiency. This is because, in the educational context, it does not make sense to speak of a ``total knowledge'' of the population since nothing prevents all individuals in a social group from having the maximum performance that can be measured. It is also not appropriate to assume a uniform distribution as the reference distribution since, when one person learns, others do not lose because of this.

The definition of the benchmark distribution of performances should be done in a contextualized manner. In the present article, which used data from the Brazilian evaluation system of basic education, a methodology based on the comparison of the performance on the PISA of Brazilian students with the performance of a typical OECD
country, was used. The Kullback-Leibler divergence was used to measure the \textit{distance} among the distributions by creating indicators of inequality and of quality. Thus, the methodology introduced in the present article can be seen as the extension of the Theil index to educational situations in which the available data is the performance of students on tests. A similar problem was dealt with by Holland (2002) \cite{holland2002two} who,  considers, however, that learning distributions can differ in only two ways, which he named horizontal and vertical. The formulation via KL resulted in an indicator that is sensitive to any kind of divergence among the distributions. Different forms of divergence among distributions require different educational policies to overcome them. The application of the methodology proposed to characterize intraschool exclusion in Brazilian municipalities showed that there are few Brazilian municipalities wherein both, the level of learning is adequate and the inequality among students of lower socioeconomic level is equivalent to that of students of higher level. 

Thus, this article presents evidences of intraschool exclusion in most the municipalities. However, unlike other studies, the present article showed the municipalities in which the reason for the exclusion is only the level, and the municipalities in which the reason is the inequality. This second reason has not yet been identified, since the indicators used in the educational debate emphasize only the level reached by the students. Thus, the present article shows clearly that Brazilian basic education has serious learning problems, and that the necessary urgent policies must seek simultaneous consideration of the problems of quality and inequality. To date, due to the lack of indicators, there is a position in the Brazilian educational debate, often latent, that it is sufficient to solve the problem of quality that the problem of inequality would be solved. The Brazilian data analyzed in the present article show that this does not occur in practice. Fortunately, public policies to elevate the level can, with simple modifications, also address the problem of inequality.



\newpage

\printbibliography

\end{document}